# Characterisation of radioactive decay series by digital autoradiography, part 1: a theoretical approach using time and space coincidence (TSC) analysis

Sardini, P.[1*] ; Donnard, J.[2] ; Descostes, M.[3,4]

1 : IC2MP – HydrASA, Poitiers University UMR 7285 CNRS, France

2 : AI4R (SAS), 2 rue Alfred Kastler, Nantes, France

3 : ORANO Mining, Environmental R&D Dept, 125 avenue de Paris, 92320 Chatillon, France

4 : Centre de Géosciences, MINES Paris, PSL Research University, Paris, France

September 2024

**Abstract**

The three natural decay chains have short-lived daughter elements, and the existence of these radioelements makes it possible for alpha and beta particle emissions to be generated at the same place and the same time. We show theoretically that such time and space coincidences (TSCs) can be detected efficiently by suitable autoradiographic systems using an algorithm that is six times more efficient than an approach based on the classical slicing of space-time. Two types of TSC coexist: true TSCs, resulting from the decay of short-lived daughter elements, and random TSCs. True TSCs are predictable and their numbers vary linearly with activity; the prediction of true $\alpha/\alpha$ and $\alpha/\alpha/\alpha$ TSCs of the $^{235}$U chain is presented. Random coincidences are also predictable using Poisson's law. They vary quadratically as a function of activity. Examination of the case of an uranium ore at secular equilibrium shows that the observed $\alpha/\alpha$ coincidences result from the sum of random and true TSCs. For high uranium contents, random coincidences predominate. For uranium at secular equilibrium, the theoretical calculation shows that true TSCs predominate for contents below ~5000 ppm.

## 1. Introduction

The characterisation of both natural and artificial radioactive disintegration series is classically performed using alpha, beta and gamma spectroscopic techniques [1,2], mass spectrometry [3], and liquid scintillation counting [4]. Mainly appropriate for bulk materials analysis, these techniques are currently employed to assess the equilibrium state of the series, and to quantify the activity of some target radionuclides within them. Sample preparation for spectroscopic or chemical characterisation plays an essential role in the analysis protocol, especially if the analysis concerns only a fraction of the material, such as for some target minerals in mine tailings [5,6].

Autoradiography can also be used to determine the 2D distribution of radioactivity in a heterogeneous sample section [7]. Applied to natural radioactive decay series, it has been demonstrated that alpha particle autoradiography can accurately locate radioactive emissions in heterogeneous samples [8,9,10,11,12,13,14,15,16]. This technique offers an interesting and non-destructive alternative to the sequential extraction method, a topic recently highlighted in the literature for detecting $^{226}$Ra [17,6]. Furthermore, a combination of autoradiography and spectroscopy, called spectroscopic autoradiography (SA), has been recently investigated for studying alpha particles in the $^{238}$U chain. It combines the local measurement of an energy distribution of alpha particles and the distribution map of alpha emissions at a larger scale [18,19]. The present contribution is focused on an alternative method for determining the activity values of some alpha and beta emitters in decay chains: the coincidence method. This technique is based on the temporal analysis of two particles collected in a detector [20,21]. For instance, consider two successive daughters A and B of a decay series, such as A→B. If the half-life of B is "short', the time lag separating the detection of the two particles emitted by A and B is "short", and thus these particles can be observed almost simultaneously thanks to the rapid disintegration of B; the temporal occurrence of such a particular event is called here a coincidence. Behind the term "short" lies a more complex reality, namely that the detectability of a coincidence depends on the total activity; for instance, if the total activity is very high, even a true coincidence due to a short-lived radioisotope will not be easily distinguished from a random coincidence. Conversely, a relatively long-lived isotope can provoke a true coincidence if the activity is low.

Some specific detectors have been developed for analysing coincidences, for example, some standard liquid scintillation counters [22]. Beta/gamma coincidences [23] are employed for detecting radioxenon using the SPALAX system [24,25,26]. The coincidence method is also employed in the RaDeCC system for estimating [223]Ra and [224]Ra activity in environmental samples [27,28,29]. In addition, in the [238]U decay chain, the β/α coincidence linked to the successive emissions of the radioisotopes [214]Bi and [214]Po has been analysed temporally [30, 31]. The coincidences described in all these works are solely temporal.

In contrast, this paper studies the analysis of particle coincidences using autoradiography: specifically, only α and β particles will be addressed here, even if γ emission could be of use, but only temporally. α/α and β/α coincidences that can be analysable on decay series are shown in Table 1. The relevant isotopes are characterised by a relatively short half-life; their massic contents are consequently very low in a sample, thus they are impossible to detect and quantify locally by chemical imaging at a micrometre scale. The analysis of such radioelements is of particular interest in the fields of biology, Earth, material and medical sciences.

Table 1 : Some coincidences involving α and/or β particles typically used for analysing decay series

| Series | Main Field | Short-lived daughter | Half-life | Type |
|---|---|---|---|---|
| [238]U | Material sciences, Medical : contamination | [214]Po | 164μs | β/α |
|  |  | [218]Po | 3.1min | α/α |
| [235]U/[227]Th/[223]Ra | Material sciences, Medical : alpha therapy | [215]Po | 1.78ms | α/α |
|  |  | [219]Rn | 3.96s | α/α |
| [232]Th/[212]Pb | Material sciences only Medical : alpha therapy | [216]Po | 140ms | α/α |
|  |  | [220]Rn | 55s | α/α |
| [237]Np/[225]Ac | Medical: alpha therapy + hopefull eight [32] | [217]At | 32ms | α/α |

The central idea of the presented work is the detection of coincidences by means of a real-time autoradiographic system, i.e. coincidences are not only detected by time delay, but also according to spatial coincidences. To our knowledge, in the field of radioactivity analysis, the concept of time and space coincidences (TSCs) was initially proposed by [33]. This paper addressed the detection of $^8$He and $^6$He ion decay products and decay half-lives. Since then, TSCs have become widely used for forming tomographic images with Positron Emission Tomography (PET). In the framework of decay chain characterisation, we now propose a novel methodology based on TSC analysis for locating short-lived daughter elements of interest in a heterogeneous sample. This new methodological approach is typically capable of locating the radioactive daughters *in situ*, which is not the case for the methods used for bulk analysis (spectroscopy, time coincidence) or for standard autoradiographic techniques such as film or phosphor screen autoradiography, which are known not to discriminate emitted particles [34].

The theoretical modalities of TSC will be described herein. TSCs differ from time coincidences by the combination of temporal and spatial information. TSCs are easily visible in the gel autoradiography provided by [13]. However, the time and space coordinates are not automatically recorded by this method, and the TSCs cannot be statistically studied. TSCs have been described for β/α coincidences ($^{214}$Bi/$^{214}$Po) in [35]. However, in that paper, coincidences were searched for in two distinct steps. Firstly, they were detected as a function of time lag, then they were projected into the section under study: thus the coincidences were not detected using a combined time and space method. In the present contribution, the second section describes the theoretical background of a new algorithm called "XYT", created specifically to find TSCs directly in a file where particles are chronologically sorted according to their emission time T, *and* also where the 2D positions of particles are respectively determined. For detecting TSCs, the XYT algorithm scans each particle with a temporal window combined with a 2D spatial window. The prediction of TSCs is developed in the third section, where random and true TSCs are evaluated separately. The dependency between total activity and the number of TSCs is provided, and also the effect of the size of the time and space windows on the detected number of TSCs. The efficiency of the XYT algorithm is pinpointed by comparing it with the Poisson point-process approach. TSCs are then evaluated for a geological sample where $^{235}$U and

$^{238}$U series are at secular equilibrium (α/α coincidences due to $^{215}$Po decays). A companion paper [36] illustrates experimentally the theoretical approach presented here.

## 2. Theorical background related to TSC analysis using the XYT algorithm

Coincidence analysis is performed by scanning an input file obtained from an autoradiographic system which is able to (1) record the emission time of each particle, (2) find the 2D emergence position (X,Y) of each particle with sufficient accuracy, and (3) to differentiate alpha and beta particles. Such systems are for instance described in [9,10,31]. The Beaquant system [37,38] is used in a companion paper [36] for testing the method developed herein. For instance, the structure of the file obtained from such autoradiographic systems is the following:

- One line represents one particle. Lines are numbered from 1 to $N_t$ ($N_t$ is the total number of particles recorded during the total acquisition time $T_{aq}$). Line indices represent the order of detection of the particles (i.e. the lines are sorted chronologically according to the detection times $T_i$); each line i contains:

(1) the detection time $T_i$ of the i-th particle (T = 0 at the beginning of the acquisition). The dead time of the acquisition device should be compatible with the half-life of the coincidence that is to be detected.

(2) the 2D coordinates ($X_i$, $Y_i$) of the emergence point of the particle.

(3) the type of particle detected (α/β). If the detected coincidences are related to a single type of particle (α/α coincidence, for instance), this information can be omitted but the acquisition system should be able to differentiate between particle types when building the output file.

From two events i and j where i < j, the Euclidian distance d and the time lag T are extracted ( $d = \sqrt{(x_j - x_i)^2 + (y_j - y_i)^2}$ and $T = T_j - T_i$). A TSC is defined by two parameters $d_w$ and $T_w$. $d_w$ is the maximal spatial distance between the two events and $T_w$ is the maximum time lag between the two events. To be classified as a "pair" the two events must respect $d \leq d_w$ and $T \leq T_w$. $d_w$ and $T_w$ are extremely important, as they represent the sizes of the space and time windows used to detect the TSCs. $d_w$ should be chosen according to the spatial resolution of the autoradiographic system used. The resolution is also variable thanks to the type of particle detected, and to its emission energy. The reader is referred to [36] for more details on the experimental procedure for determining $d_w$. $T_w$ can be determined as a function of the half-life of the emitter responsible for the coincidence (10 times the half-life can be used as an upper limit).

The XYT algorithm is quite simple to implement. It is basically described by the successive scanning of the autoradiographic file, containing $N_t$ lines (particles). i represents the current scanned particle during the process ($1 \leq i \leq N_t$). The algorithm starts at the first particle ($i = 1$), and scans each subsequent line of the file one by one, for $j > i$. The scan continues while $T_j - T_i < T_w$, except if the distance between particles i and j is lower than $d_w$. If this exception does happen, the particles i and j are found to be in time and space coincidence, and are tagged into a vector, and the scan is stopped. If this exception does not happen, the particle i is not tagged, and the process continues as before to the next untagged particles of the file. The next i index, from which the search resumes is i+1. If index i+1 is tagged, the algorithm continue the research forward (i+2), etc. We have not consider the possibility of multiple TCS pairs here. This would be possible to investigate by searching in the time window all possible pairs respecting the space criterion. At the moment, the first pair found in the time window is the one which is selected. In case of coincidences by two different particle types, care should be taken that the algorithm only selects start (i) and stop (j) events corresponding to the appropriate particle type. Using this algorithm, each particle of the file is scanned and possible TSCs are detected. The algorithm is also illustrated by a programming flowchart shown in Figure 1.

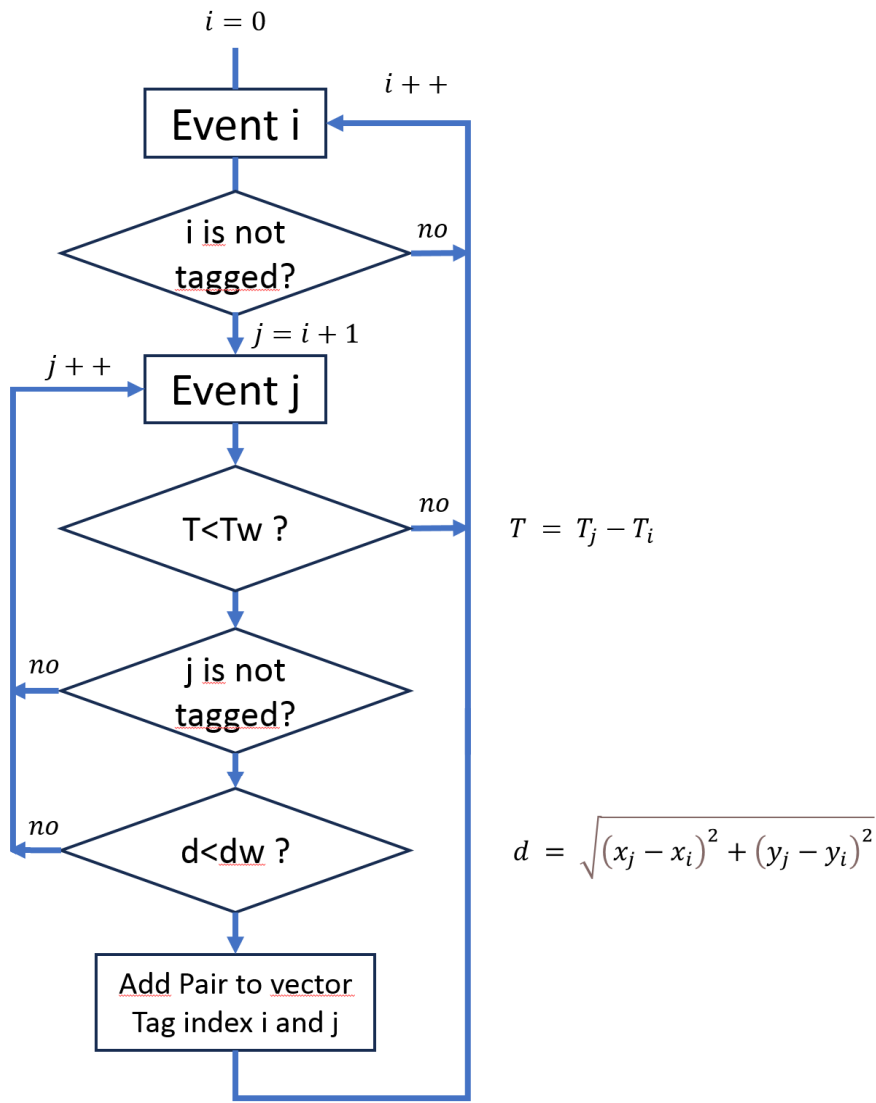

Figure 1 : Flowchart of the XYT algorithm. This flowchart corresponds to the case where all particles are of the same type (no test on the type of particle).

## 3. Predictions from TSCs

Irrespective of the particle type, two types of TSC will be found to coexist when analysing radioactive samples. True coincidences are created by particle emissions which are associated with the creation and decay of the same nucleus. In case of random coincidences, one should emphasize that they arise from uncorrelated decay events (from unrelated nuclei). If we consider a single TSC detected by the XYT algorithm, it is not possible to know from which of these two processes it originated. However, by considering a set of detected coincidences, it is possible to predict the proportion of the two types of TSC. The numbers of true and random TSCs are predicted for the activities found in "natural" geo-materials. The maximum alpha counting rate considered $n_{max}$ = 2 cps/mm² is close to that of uraninite at secular equilibrium [10]. This emission rate is equivalent to 55.7 U wt%.

**3.1 Random Coincidences**

*3.1.1 Double coincidence of a Poisson point process*

If the time and space projection of the particles recorded by the autoradiograph is purely random, then it can be considered as a spatialized Poisson point process. In such a case, the distribution of the particles in the XYT space follows the rule of a Poisson distribution.

When considering a hypercube of dimension $S_{aq}*T_{aq}$, partitioned by a regular grid ($S_{aq}$ is the 2D area of the autoradiograph, and $T_{aq}$ is the total acquisition time), the probability of finding y points (particles) (y = 0, 1, 2, …) in any element of the grid is defined by:

$$P(Y = y) = \frac{\lambda^y}{y!}e^{-\lambda}, \quad y = 0, 1, 2, \ldots \tag{1}$$

where λ [0; +∞] is the Poisson parameter, and Y is the number of points projected in any element of the grid.

λ is the mathematical expectation of the Poisson law, i.e. the total number of particles detected $N_t$, divided by the number of elements composing the grid $N_g$:

$$\lambda = N_t/N_g \text{ with } N_g = (S_{aq} \times T_{aq})/(d_w^2 \times T_w) \tag{2}$$

$d_w$ [L] and $T_w$ [T] are the size (in XY space) and the time defining the elementary dimension of the grid elements, respectively.

The number of double-random coincidences $N_{rco}$ occurring in these $N_g$ hyper-volumes is determined by:

$$N_{rco} = N_g \times P(Y = 2) = N_g \times \lambda^2 / 2 \times \exp(-\lambda) \qquad (3)$$

The P(Y=2) case represents the probability of an exact double coincidence (only two particles are detected), while P(Y>1) is the probability that at least two particles are detected in coincidence. So, in the latter, it is also allowed that more than two particles are detected within the coincidence window. Since in equation (3), occurrences of three or more coincidences are ignored, it would have been more accurate to use P(Y>1) instead of P(Y=2). Nevertheless, it has been verified that the difference between P(Y>1) and P(Y=2) is negligible (close to 0.2% in the worst case, i.e. for $n_{max}$).

Now, let us consider the case of time and space autoradiography where ($d_w^2$, $T_w$) are far smaller than ($S_{aq}$, $T_{aq}$). For instance, consider first the case where $S_{aq}$ = 1 cm$^2$, $T_{aq}$ = 1 day, $d_w$ = 413 µm and $T_w$ = 17.8 ms. The chosen values for $S_{aq}$ and $T_{aq}$ are consistent with the conditions routinely employed for geo-material characterisation. The choice of $d_w$ is related to the experimental spatial resolution of the autoradiographic device determined in the companion paper [36]. The choice of $T_w$ is 10 x $T_{1/2}$ of $^{215}$Po (1.78 ms). Using these parameters, $N_g$ = 2.84 x 10$^9$ and $N_t$ = 1.73 x 10$^7$ using $A_{max}$. Then, for that case, $\lambda$ is close to 0 ($\lambda$ = 6.07 x 10$^{-3}$), and thus equation (3) can be simplified to:

$$N_{rco} = N_g \times \lambda^2 / 2 \qquad [4]$$

Combining equations (2) and (4), and defining the total counting rate n (cps/mm$^2$) as $N_t/(S_{aq} \times T_{aq})$, yields:

$$N_{rco} = n^2/2 \times (d_w^2 \times T_w) \times (S_{aq} \times T_{aq}) \qquad (5)$$

Equation [5] predicts that the number of random TSCs is proportional to the square of the total counting rate n, to the size of the elementary grid element ($d_w^2$ x $T_w$), to the acquisition time $T_{aq}$ and to the whole autoradiograph area $S_{aq}$. Dividing each side of equation (5) by $S_{aq}$ x $T_{aq}$ provides the counting rate of random coincidences $n_{rco}$ (cps/mm$^2$):

$$n_{rco} = n^2/2 \times d_w^2 \times T_w \qquad (6)$$

### 3.1.2 Comparison with the XYT algorithm

$d_w^2$ and $T_w$ are the two parameters defining the size of the time and space window employed to count the random TSCs using the XYT algorithm. The number of TSCs it obtains is however expected to be different from the one obtained from Poisson's law's prediction (Figure 2). Indeed, because spatial partitioning using the XYT method is adaptative by searching the nearest neighbour particle, the XYT method is expected to be more efficient than prediction using a Poisson distribution; a significant number of coincidences are ignored using a regular time and space partition, which is illustrated in Figure 2.

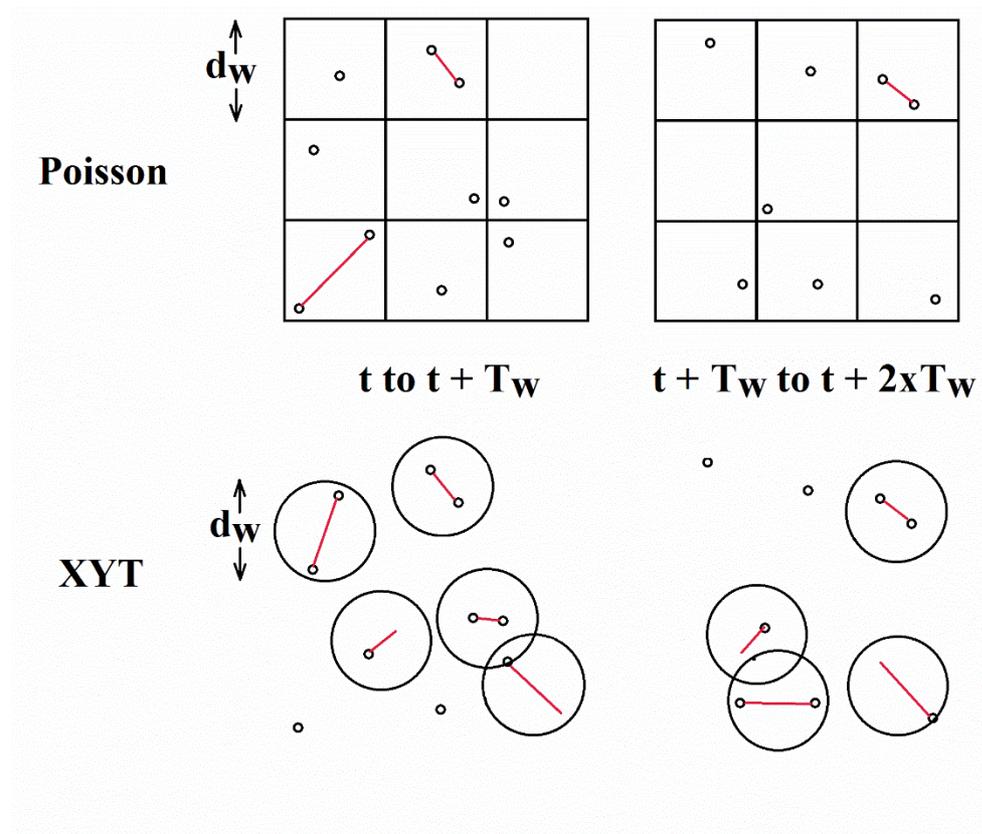

Figure 2: Possibilities of coincidence detection using either regular partitioning or the XYT algorithm. These cases illustrate that the number of coincidences found by regular partitioning is much lower than those found by the XYT algorithm.

Simulations of random TSCs were performed using 'R' statistical software. Then, the XYT algorithm was used to study the relationship linking the counting rate of random TSCs $n_{rco}$ to the total counting rate n. Between 100 and 500,000 particles were projected randomly and uniformly into the XYT space, always using $S_{aq}$ = 1 cm$^2$, and $T_{aq}$ = 1 h, and also using repetitions (for instance for the simulation of 100 random particles projection, we employed 50,000 repetitions to determine $n_{rco}$). For these calculations, the imposed emission rate ranged from 2.78 x 10$^{-4}$ to 1.39 cps/mm$^2$.

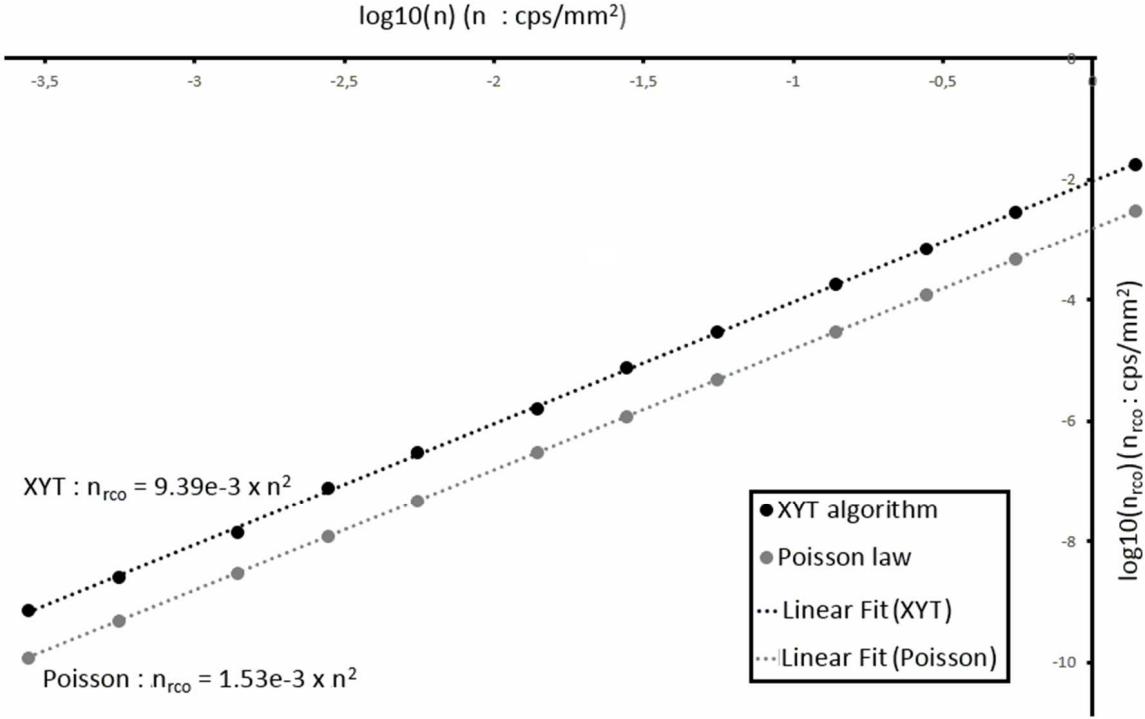

Figure 3 : $n_{rco}$ as a function of total counting rate n, obtained by the XYT algorithm and by Poisson's law. $d_w$ = 413 µm and $T_w$ = 17.8 ms.

For both Poisson's law and the XYT algorithm, $n_{rco}$ is plotted as a function of n (Figure 3). It was checked that the relationship between $n_{rco}$ and n follows a quadratic relation, a result expected for both methods (see equation (6)). Because the relationship was checked in the range of activity encompassing the alpha radioactivity range found in the U decay series in the environment, the approximation taken for transforming equation (3) to (4) is validated. Furthermore, it was found that the XYT algorithm is about 6.12 times more efficient than the Poisson method using regular partitioning. It was also assessed that varying the size of the time and space window for the XYT algorithm has the same effect as predicted by equation (6). Note that the number of random TSCs obtained by the XYT algorithm can be predicted not only according to the total counting rate of the studied section, but also according to the size of the time and space window which is used to detect them. When using the XYT algorithm, the following equation can be employed to estimate $n_{rco}$ (cps/mm²) in relation to $n$ (cps/mm²), by introducing a correction factor $\xi$:

$$n_{rco} = \xi \times n^2 \times d_w^2 \times T_w, \text{ with } \xi = 3.06 \qquad (7)$$

In practice, in order to minimize the detection of random TSCs, equation (7) shows that it is important to optimize the size of the time and space window (Figures 4a, 4b). Indeed, because random TSCs represent noise, it is recommended that its size be minimised.

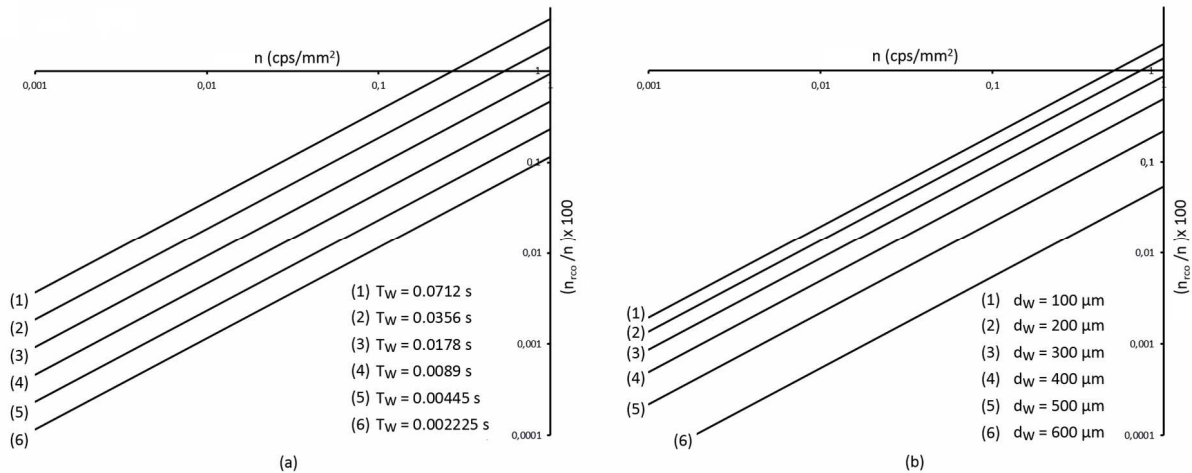

Figure 4: Counting rate of random double coincidences ($n_{rco}$) is plotted as a function of total counting rate n, (a) for different values of $T_w$, and (b) for different values of $d_w$.

### 3.1.3 Comparison between TSC and only time coincidences

The classical coincidence method is based only on time measurements performed on bulk samples (solid or liquid). However, for the same counting rate, TSCs differ from random time coincidences. The lambda parameter of Poisson's law ($\lambda = N_t/N_g$) depends on $N_g$, which is merely equal to $T_{aq}/T_w$ for temporal analysis. In this case, $\lambda$ cannot be considered to be close to 0 for any emission rate found in geological samples; for instance, using $T_{aq}$ = 1 day and $T_w$ = 17.8 ms, $\lambda$ = 3.56 for an emission rate of 2 cps/mm$^2$. Table 2 compares the number of random coincidences obtained either by time and space or temporal analysis.

Table 2: Number of random coincidences $N_{rco}(Y>1)$ determined by Poisson's law, for temporal and time and space coincidence methods ($T_{aq}$ = 1 day, $T_w$ = 17.8 ms, $S_{aq}$ = 1 cm$^2$ and $d_w$ = 413 µm). $N_t$ ranges from $N_{max}$ = 1.73 x 10$^7$ to $N_{max}$/10000 particles.

| $N_t$ | 1.73 x 10$^7$ | 1.73 x 10$^6$ | 1.73 x 10$^5$ | 1.73 x 10$^4$ | 1.73 x 10$^3$ |
|---|---|---|---|---|---|
| $N_{rco}(T)$ temporal coincidences | 4.22 x 10$^6$ | 2.43 x 10$^5$ | 3004 | 31 | 0.31 |
| $N_{rco}(ST)$ TSC | 52,252 | 524 | 5.24 | 5.24 x 10$^{-2}$ | 5.24 x 10$^{-4}$ |

It appears that the number of random coincidences is much higher when using only time methods, especially for low activities: the ratio $N_{rco}(T)/N_{rco}(TS)$ is higher than 500 for counting rate lower than 0.1 cps/mm$^2$. This represents an important benefit of using TSC instead of solely temporal coincidences. This difference comes mainly from the fact that the number of intervals tested ($N_g$) is much higher (almost three orders of magnitude) in the case of time and space, because the discretisation in the XY dimension is not present for the temporal case. For instance, using the parameters of Table 2, $N_g$ = 4.85 x 10$^6$ compared with 2.85 x 10$^9$, for temporal and TSC approaches, respectively. These results are not surprising, true coincidences always occuring localized because they are associated with the decay of the same nucleus. On the other hand, random coincidences are more likely not to be localized in the used spatial window, because the particle emission originates from the decay of unrelated nuclei, which can be in different parts of the sample. A solely time-based analysis consider particle emission from any part of the sample, whereas TCS places a constrain on the effective size of the area considered (i.e. the size of the spatial window).

## 3.2 True coincidences

In a given sample, true coincidences originate from short half-life emitters in their decay series. Considering such radionuclides (RNs), a first particle is emitted at the same time as the short half-life RN is created. After a short time lag, this first emission is followed by a second emission due to the disintegration of the short half-life RN: that is a true coincidence. The next challenge is to predict the number of true TSCs observed at the sample surface according to the bulk activity. Let us consider the simplest case, that of alpha particles: when they are emitted by a disintegration, their trajectory is almost straight [39]. Furthermore they are emitted at a single emission energy $E_0$ for $^{238}U$ and $^{235}U$ series. Consider the emission of a single alpha at a depth L below the sample surface, and define $P_{OUT}(L)$ and $P_{IN}(L)$ as the probabilities of observing, or not, this particle emerging at the sample surface, respectively. These two probabilities can be written according to the emission depth as [40]:

$$P_{OUT}(L) = 0.5 \times \left(1 - \frac{L}{Rmax}\right) \quad (8)$$

$$P_{IN}(L) = 0.5 \times \left(1 + \frac{L}{Rmax}\right) \quad (9),$$

where $R_{max}$ represents the maximum range of alpha emissions in the sample. Average values of $P_{OUT}(L)$ and $P_{IN}(L)$ for different types of events are given in table 3. The first type described corresponds to the emission of a single particle at depth; the average probability of emergence is 1/4 [41]. The second type corresponds to the emission of two alpha particles at depth, nearly at the same location and at the same moment: this is an α/α coincidence. The average probability $P_{OUT}$ to observe an α/α coincidence at the sample surface is 1/12 (table 3). For triple alpha coincidences, $P_{OUT}$ is 1/32.

Table 3. First column: illustration of alpha emissions within sample thickness (IN) or towards the detector (OUT). Second column: integrals used for the calculation of average probabilities P; indexes i and o refer to "in" and "out" respectively; P values are given in the third column.

| One alpha emitted | Average probability of observation P | P value |
|---|---|---|
| 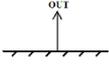 OUT | $P_o = \dfrac{1}{R_{max}} \displaystyle\int_0^{Rmax} P_{OUT}(L)\,dL$ | 1/4 |
| 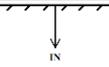 IN | $P_o = \dfrac{1}{R_{max}} \displaystyle\int_0^{Rmax} P_{IN}(L)\,dL$ | 3/4 |
| **Two alphas emitted** | **Average probability of observation P** | |
| 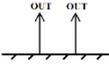 OUT OUT | $P_{oo} = \dfrac{1}{R_{max}} \displaystyle\int_0^{Rmax} P_{OUT}^{\,2}(L)\,dL$ | 1/12 |
| 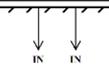 IN IN | $P_{ii} = \dfrac{1}{R_{max}} \displaystyle\int_0^{Rmax} P_{IN}^{\,2}(L)\,dL$ | 7/12 |
| 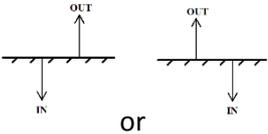 OUT OUT / IN IN or | $P_{io} = P_{oi} = \dfrac{1}{R_{max}} \displaystyle\int_0^{Rmax} P_{IN}(L) P_{OUT}(L)\,dL$ | 2/12 |
| **Three alphas emitted** | **Average probability of emission P** | |
| 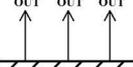 OUT OUT OUT | $P_{ooo} = \dfrac{1}{R_{max}} \displaystyle\int_0^{Rmax} P_{OUT}^{\,3}(L)\,dL$ | 1/32 |

Note that the examples chosen here only cover those coincidences when alpha particles are emitted. However, beta and alpha emissions coexist in decay chains. The prediction of TSCs involving beta particles (β/α, α/β, and β/β) is not developed here because of the complexity inherent in beta emission. This aspect is discussed in the last section of this paper.

Let us now consider the succession of three alpha disintegrations involving four RNs of the decay chain A→B→C→D at secular equilibrium (the alpha emission points are bounded by the 0 and $R_{max}$ planes). B and C are the short-life RNs; their rapid disintegration is able to generate observable α/α and α/α/α TSCs. In other words, the approach described below is not valid for long-life RNs. For instance, for the $^{235}$U series, the case $T_{1/2}(A) \gg T_{1/2}(B) \gg T_{1/2}(C)$ corresponds to A = $^{223}$Ra ($T_{1/2}$ = 11.3 d), B = $^{219}$Rn ($T_{1/2}$ = 3.96 s) and C = $^{215}$Po ($T_{1/2}$ = 1.78 ms). Let $n_\alpha$ (cps.mm$^{-2}$) be the counting rate of α particles detected for one disintegration reaction *alone* (C→D for instance). Then, from $n_\alpha$ it is possible to go back to the TSCs using the following approach:

- 4 x $n_\alpha$ is the activity of alpha particles emitted in the solid sample for C→D decay alone (disintegration of $^{215}$Po)
- 8 x $n_\alpha$ is the activity of alpha particles emitted in the solid sample for B→C→D decays (disintegration of $^{219}$Rn and $^{215}$Po)
- $n_{tco}$ = 8 x $n_\alpha$/12 = 2 x $n_\alpha$/3 is the counting rate of true α/α coincidences theoretically observable for B→C→D decays, using the average probability of observation at the sample surface of a double TSC (1/12) for two successive disintegrations (Table 3). These double TSCs involve 4 x $N_\alpha$/3 alpha particles.
- $n_{tco}$ = 12 x $n_\alpha$/32 = 3 x $n_\alpha$/8 is the counting rate of true α/α/α TSCs theoretically observable for A→B→C→D decays, using the average probability of observation at the sample surface of a triple TSC (1/32) for three successive disintegrations (Table 3).

## 3.3. Prediction of α/α TSCs for $^{219}$Rn/$^{215}$Po emissions ($^{235}$U decay series)

In a decay series, the TSCs observed in a section are either true or random. Random coincidences originate from long-life RNs, but also partly from short-life RNs. The true coincidences originate exclusively from short-half-life RNs.

We focus here on a first example: the $^{235}$U series, and the coincidences of α/α type coming from $^{219}$Rn and $^{215}$Po successive decays, for a geological sample where $^{235}$U and $^{238}$U series are at secular equilibrium. The true TSCs come from the activity of the $^{235}$U decay series, and counting rate $n_{tco}$ is only a function of this activity referred to as

$n_{235}$. Thus, the total counting rate coming from these true TSCs is $n_{tco} = 2 \times n_\alpha/3$, where $n_\alpha$ is 1/7 of $n_{235}$.

However, the random TSCs correspond to the activity of both $^{235}U$ and $^{238}U$ decay series. Indeed, in the case of a geological sample containing U at secular equilibrium, the two chains coexist. Note however that the activity of $^{235}U$ series can be neglected compared to the activity of $^{238}U$ series because the activity of $A_{235}$ is only around $\varepsilon$ = 3.7% of the total alpha activity of the $^{238}U + ^{235}U$ series. If equation (7) is simplified using $d_w$ = 413 µm and $T_w$ = 17.8 ms, it yields $n_{rco} = \xi' \times n^2$ ($\xi'$ = 9.3 x $10^{-3}$ mm².s). Then, the total counting rate of α/α coincidence $n_{co}$ ($s^{-1}.mm^{-2}$) is $n_{co} = n_{tco} + n_{rco}$, with:

$$n_{tco} = 2 \times n_{235}/21 = 2 \times \varepsilon \times n/21 \quad \text{and} \quad n_{rco} = \xi' \times (n - 2 \times n_{tco})^2 \quad (10)$$

We have finally:

$$n_{co} = 2 \times \varepsilon \times n/21 + \xi' \times (1 - 4 \times \varepsilon / 21)^2 \times n^2 \quad (11)$$

In (11), the term $(1 - 4 \times \varepsilon \times / 21)^2$ = 0.986 represents the fact that in (10), the alpha emission rate contributing to the true TSCs should be deducted from the total counting rate when random TSCs are calculated. However, this term being very close to 1, it can be neglected in a first approximation in (11).

Equation (11) is interesting because it shows that the number of TSCs found in a given surface of counting rate n is a simple function of the number of true and random TSCs. If this surface can be subdivided into a set of 'regions of interest' (ROIs), the local counting rate of TSCs determined in each ROI plot according to the related counting rate of the ROI would follow (11). This is however only valid if the equilibrium state in the section is homogeneous. According to equation (11), there would be a possibility of separating true and random TSCs by, for instance, calculating the first-order derivative $dn_{co}/dn$ at the origin of a (n, $n_{co}$) scatter plot. This calculation would determine the number of true TSCs. Conversely, if the scatter plot (n, $n_{co}$) did not follow this law, the equilibrium state of the section would be more complex. Finally, Table 4 shows some values of $n_{tco}$, $n_{rco}$ and $n_{co}$ for total counting rates n ranging from 2 to 2 x $10^{-5}$ cps/mm². As a result, it is clear that true TSCs will be predominant for emission rates lower than about 2 x $10^{-2}$ cps/mm² (around 5000 ppm U).

Table 4. n is the total counting rate, and $n_{rco}$ and $n_{tco}$ are counting rates of random and true α/α coincidences, respectively, calculated with [10]. All counting rates are given in cps/mm$^2$. $^{235}$U and $^{238}$U decay series are at secular equilibrium. The conversion in U (wt%) is based on the alpha activity of pure uraninite (U wt% is taken at 80%, giving n = 2.5 cps/mm$^2$).

| n | 2 | $2 \times 10^{-1}$ | $2 \times 10^{-2}$ | $2 \times 10^{-3}$ | $2 \times 10^{-4}$ | $2 \times 10^{-5}$ |
|---|---|---|---|---|---|---|
| U ppm | 557,000 | 55,700 | 5570 | 557 | 55.7 | 5.57 |
| $n_{tco}$ | 7.05E-04 | 7.05E-05 | 7.05E-06 | 7.05E-07 | 7.05E-08 | 7.05E-09 |
| $n_{rco}$ | 3.67E-02 | 3.67E-04 | 3.67E-06 | 3.67E-08 | 3.67E-10 | 3.67E-12 |
| $n_{co}$ | 3.74E-02 | 4.37E-04 | 1.07E-05 | 7.41E-07 | 7.08E-08 | 7.05E-09 |
| $n_{tco}/n_{co}$ % | 1.89 | 16.12 | 65.77 | 95.05 | 99.48 | 99.95 |

## 4. Discussion/Conclusion

This paper presents some predictions of α and/or β particle emissions in natural decay series, generating TSCs that are measurable both in 2D space and time dimensions. This contribution defines the theoretical framework for a new methodology enabling the detection of double or triple coincidences, opening up the possibility of precisely locating certain radioelements of natural decay chains present in geological materials. For detecting TSCs, a dedicated algorithm named "XYT" was developed and tested. Its ability to find coincidences was analysed by comparing it with Poisson's law's predictions, for the case of random coincidences. The XYT algorithm was found to be significantly more efficient than the statistical approach using Poisson's law. It was demonstrated that the dependency of the amount of detected random coincidences with total counting rate is impacted by the size of the spatial and temporal windows ($d_w$ and $T_w$). For an optimal detection of true coincidences in geo-materials, these findings emphasise the importance of choosing these two parameters carefully. Because the number of detected random coincidences is a linear function of $d_w^2$ and $T_w$, one should minimise these two parameters so as to decrease the number of random coincidences. For instance if a time window of $T_w = 10 \times T_{1/2}$ is used, the probability of emission is almost 100% in this window (1-1/2$^{10}$ = 99.9 %). A time window of eight periods would perhaps be more suitable, because emission probability within this window would be near 99.6%, but the number of random coincidences would be four times less than using 10 x $T_{1/2}$. The same type of reasoning is applicable to the size of the spatial window. The choice of $d_w$ is explained in [36], because it is determined experimentally. However, the predictions of random coincidences developed in this paper remains to be adapted to heterogeneous materials. Indeed, these predictions have been studied herein using spatially-uniform random distributions. However, this assumption of uniformity of distribution is very rarely verified, as activity is often distributed in the form of hot spots, embedded in a continuous background [12,15,6]. One solution would be to subdivide the study area into ROIs of uniform counting rate, and to determine the counting rate of random coincidences in each ROI.

In the future, the theoretical model of TSC should be extended to take into account the detection efficiency, and thus provide more accurate predictions for the random and true coincidence counting rates. For example, for alpha particle detection, the Beaquant gas detector has an efficiency of around 80% [10].

This work gives the expression of true coincidences only for those involving $\alpha$ particles (double or triple coincidences). An example of such coincidences are those in the $^{235}$U chain, involving the succession $^{223}$Ra, $^{219}$Rn and $^{215}$Po. In this contribution, coincidences involving $\beta$ particles have not been formally studied. To do this, it will be necessary to determine the probability functions for the emergence of beta particles at the sample surface. These functions, named above $P_{IN}(L)$ and $P_{OUT}(L)$, depend on the depth L of the emission point. They can only be determined by simulation, as the trajectory of $\beta$ particles in matter is not rectilinear, and $\beta$ particle emission energy is distributed.

The case of $\alpha/\alpha$ TSCs due to the rapid decay of $^{215}$Po for a U ore at secular equilibrium has been examined; generally, the total TSCs result from the sum of random and true TSCs (see equation (10)). This equation is interesting for two reasons. Firstly, it demonstrates that it is possible to find the contribution of both types of coincidences on a plot diagram of total counting rate vs. counting rate of double coincidences. On such a plot diagram, each point could represent a different ROI. Secondly, if several equilibrium states were spatially distributed on the same section, it would be possible to separate them, since the relationship linking the number of true coincidences with the total number would differ. Of course, these conclusions remain to be confirmed experimentally.

The technique can be applied in some other cases, for instance to discriminate between artificial and natural radionuclides using autoradiography. In [42], the detection of an artificial radionuclide (plutonium) is performed on the filter paper of a dust monitor that collects dust in the air and simultaneously measures radiation. However, the detected alpha particles can also originate from natural uranium- and thorium-series nuclides. If these series are in secular equilibrium, a particle containing these radionuclides will be easily distinguished by the presence of true TSCs.

TSC analysis is an alternative to the latest developments of autoradiographic measurements dedicated to mapping the equilibrium state of $^{238}$U decay series, by combining electronic autoradiography with chemical mapping of U (using SEM EDS or WDS) [11]. Autoradiographic spectroscopy of alpha particles can be applied if the local emission rates measured in a region of interest (ROI) are sufficiently high (2000 hits is a minimal counting rate required per ROI [18,19]).

Finally, verify experimentally the existence of TSCs and the quantitative nature of their prediction is the main purpose of the companion paper [36]: this paper applies the XYT algorithm to experimentally analyse TSCs in geological thin-sections using a suitable autoradiographic apparatus. The information needed to design equipment that applies the theory of TCSs is summarized as follows:

- The ability to collect the time of each successively emitted particle, as described by [42] as a real-time autoradiography system.

- Have a dead time of at least 50 µs, in order to detect α/α coincidences due to the decay of $^{215}$Po.

- Have the highest possible spatial resolution. To minimise the detection of random TSCs, a high-resolution instrument will optimise the size of the spatial window for detecting TSCs. A resolution of less than 20 µm for the detection of a single alpha particle has been described by [42]. The distance between two particles emitted by the same nuclei can theoretically be twice the range of the particle in matter (nearly 60 µm). Thus, using the system given by [42], all TSCs involving alpha particles should be detectable using a spatial window of nearly $d_w$ = 60 + 2 x 20 = 100 µm.

- The detector must have the best possible detection efficiency.

- The system must be able to differentiate alpha and beta particles.

- Optionally, the system should be able to detect the γ emissions, for the study of temporal coincidences associated with α or β emissions.

- Optionally, for alpha coincidences, the system must be able to determine the direction of the emission trajectory of each alpha particle. Knowledge of this criterion will make it possible to exclude certain random coincidences.
- Eventually, a large field of view is of interest for autoradiography of decimetric samples.